\documentstyle[12pt,twoside,fleqn,espcrc1]{article}

\newcommand{\beq}{\begin{eqnarray}}
\newcommand{\eeq}{\end{eqnarray}}

\title{Vector Mesons in Nuclear Medium}

\author{Tetsuo Hatsuda and Hiroyuki Shiomi
\address{Institute of Physics, University of Tsukuba,
 Tsukuba, Ibaraki 305, Japan}}

\begin{document}

\maketitle

\begin{abstract}

We  summarize the current theoretical and experimental
 status of the spectral change of the vector mesons in dense
matter.

\end{abstract}

\section{Introduction}

 Effective masses of the vector mesons
 ($\rho$, $\omega$, $\phi$) in hadronic medium
 have recently attracted wide  interests.
 The decrease of the masses can be  interpreted as
 an evidence of the  partial restoration of chiral symmetry and
 as a precursor of the transition to the quark gluon plasma
\cite{review}.
 Current SPS (CERN) and AGS (BNL) experiments are
 suitable to look for such finite density phenomena
 while RHIC and LHC will serve as  machines
 for the ``hot'' quark-gluon plasma.
 Several experiments are also  planned to detect the
 partial restoration of chiral symmetry in  heavy nuclei
 through the reactions such as $\gamma+A \rightarrow A^* + e^+
e^-$ and
$p+A \rightarrow  A^* + e^+ e^- $ \cite{SE}.

In this report, we will summarize the current theoretical and
experimental
 status of the spectral change of the vector mesons in dense
matter.

\section{Partial restoration of chiral symmetry at finite
density}

 The hadronic mass shift as well as the change of the chiral
condensate
 are known to occur only when $T$ is close to  $T_c$
\cite{review}.
 On the other hand,
at finite baryon density,  one may
 expect partial restoration of chiral symmetry even
 in the heavy nuclei \cite{DL,HHP};
\begin{eqnarray}
\label{condrho}
{\langle \bar{u}u \rangle_{\rho} \over \langle \bar{u}u
\rangle_0}
 = 1- {4\Sigma_{\pi N}\over f_{\pi}^2m_{\pi}^2}
 \int^{p_{_F}} {d^3p \over (2\pi)^3} {m_N \over E_p} ,
\end{eqnarray}
where $E_p \equiv \sqrt{p^2 + m_N^2}$, $\Sigma_{\pi N} = (45 \pm
10) $MeV,
 and the integration for $p$ should be taken
  from 0 to the
 fermi momentum  $p_{_F}$.
 At normal nuclear matter density ($\rho=\rho_0=0.17/{\rm
fm}^3$),
 the above formula gives (34$\pm$8)\% reduction of the
 chiral condensate from the vacuum value.
   Several estimates show that
 corrections to this simple fermi-gas approximation
 are small at $\rho_0$ \cite{Ko94}.
 If this is the case, the heavy ion collisions and also
 the reactions in heavy nuclei are  good testing ground of
 the partial chiral restoration.

\subsection{Vector mesons in nuclear matter}

Let's
 consider $\rho$ and $\omega$ mesons propagating inside the
nuclear matter.
 Adopting the same fermi-gas approximation with (\ref{condrho})
and taking
 the vector meson at rest (${\bf q}=0$),
 one can generally write the mass-squared shift  as
\begin{eqnarray}
\label{massshiftf}
\delta m^2_{_V} \equiv m^{*2}_{_V} - m^2_{_V} = 4 \int^{p_f}
  {d^3 p \over (2 \pi)^3  }  {m_{_N} \over E_p}  f_{VN}({\bf p}),
\end{eqnarray}
where $f_{VN}({\bf p})$ denotes the vector-meson (V) -- nucleon
(N)
 forward scattering amplitude
 in the relativistic normalization.
 Here, we took spin-isospin average for the nucleon states
 in $f_{VN}$.
 If one can calculate $f_{VN}({\bf p})$ reasonably well in the
range
 $0 < p < p_{_F}=270$ MeV
 (or $1709\ {\rm MeV} < \sqrt{s} < 1726\ {\rm MeV} $ in terms of
the
 $V-N$ invariant mass), one can predict the mass shift.
 Unfortunately, this is a formidable task:
  $f_{VN}({\bf p})$ is not  constant at all
  in the above range since there are at least two
 s-channel resonances $N(1710), N(1720)$ in the above interval
 and two nearby resonances $N(1700)$ and $\Delta(1700)$.
  They all couple to the $\rho-N$ system \cite{PDG}
 and give a rapid variation
 of $f_{VN}({\bf p})$ as a function of $p$.
 Thus one should look for totally different approach to estimate
 $\delta m_V$, one of which is the QCD sum rules in medium.

\subsection{Constraints from QCD sum rules}

Starting from the retarded correlation of the vector currents
 in nuclear medium, one can write down QCD sum rules for the
 spectral functions in medium \cite{HL}.
 Parametrizing the spectral function by
 the peak position of the resonance,
 the continuum threshold and the integrated strength of the
resonance,
  one can extract the mass shift.
 In Fig. 2(a), results of  such analysis using the Borel sum rule
 are shown  \cite{HL}:
\begin{eqnarray}
{m_{_V}^* \over m_{_V}} \simeq 1 - c_{_V} {\rho \over \rho_0} ,
\end{eqnarray}
where $c_{\rho, \omega} = 0.18 \pm 0.05$ and
 $c_{\phi} = (0.15 \pm 0.05)  y$ with $y = 0.1 - 0.2$ being the
 OZI breaking parameter in the nucleon.
  $c_{_V}$ is  obtained by neglecting the contribution of the
  quark-gluon mixed operator with twist 4; inclusion of them
  moves the central value of $c_{\rho,\omega}$ to $0.15$
\cite{HLS}.

\begin{figure}[t]
   \vspace{12pc}
 \caption{(a) Masses of $\rho$, $\omega$ and $\phi$ mesons
 in nuclear matter predicted in the QCD sum rules  (left) [7]
 together
 with the prediction of the Walecka model (right) [9]. $M^*/M$
 in the right figure shows the effective mass of the nucleon.}
\end{figure}

\subsection{Use and misuse of the QCD sum rules in nuclear
medium}

  As we  have mentioned, $f_{VN}({\bf p})$ must be  a rapidly
varying
 function of $p$.
  In terms of the mass shift,
 the (invalid) approximation $f_{VN}({\bf p}) = f_{VN}(0)$
 for $0 < p < p_F$
 implies that $\delta m^2_{_V} \simeq f_{VN}(0) \rho $.
 Although this formula is valid at extremely low density,
 it is {\em useless} at nuclear matter density.
 Motivated by the formula, however,
 it is claimed 	in ref.\cite{Koike} that the mass shift is
positive.
  It can be shown that  this claim is erroneous \cite{HLS}:
Firstly,
 the mass shift and the scattering length does not have
 direct connection in nuclear matter due to the momentum
 dependence of $f_{VN}({\bf p})$.
  Secondly, sum rules for the $V-N$ scattering amplitude cannot
predict
 the $V-N$ scattering length without dimension 8 operators in
OPE.
 Thirdly, sum rules for $\omega^2 \Pi^V(\omega)$ which is adopted
 in ref.\cite{Koike}
 does not work at all even in the vacuum without
 dimension 8 operators  and so does in the medium.

\subsection{Effective theory approaches}

 There exist many attempts so far to calculate the
 spectral change of the vector mesons in effective theories.
 Chin \cite{chin},
 using the Walecka model,
 predicted the increasing $\omega$-meson mass
 in medium due to the scattering process
$ \omega + N \rightarrow
 N \rightarrow \omega + N $.
  More sophisticated calculations for  the $\rho$-meson
  predict also a slight increase of the $\rho$-mass
\cite{herman}.
  In all these calculations,  the effect
 of the polarization of the Fermi sea is only taken into account.

  On the other hand, Kurasawa and Suzuki have found
  that the mass of the $\omega$-meson is affected substantially
 by the vacuum polarization in medium
 $ \omega \rightarrow N_* \bar{N}_* \rightarrow \omega$,
 where  $N_*$ is the nucleon
  in nuclear medium \cite{kusu}. The vacuum polarization
dominates over the
 Fermi-sea polarization and leads decreasing
 vector meson mass. This conclusion was confirmed later
 by other authors \cite{others}
 and also generalized to the $\rho$-meson \cite{HS} (see
Fig.2(b)).

 What is missing in the {\em Fermi sea} approaches
\cite{chin,herman}
  is the
 effect of the scalar mean-field on the vector meson mass.
 On the other hand,  {\em Dirac sea} approaches
\cite{kusu,others,HS}
  have close similarity with other
 mean-field models
 such as those of Brown and Rho \cite{BR},
 Jaminon and Ripka \cite{JR}, and Saito and Thomas \cite{ST},
 which predict the decrease of the vector-meson masses.
   It is desirable to develop a unified  effective lagrangian
 which embodies the essential part of these approaches
\cite{KH95}.

\section{Experimental status}

 There exist already two proposals to look for the mass shift of
the
 vector mesons in nuclear medium  \cite{SE}.
 One is by Shimizu et al. who  propose an experiment
 to  create $\rho$ and $\omega$
 in heavy nuclei using coherent photon - nucleus reaction and
  subsequently  detect the lepton pairs from $\rho$ and $\omega$.
 Enyo et al. propose to create $\phi$ meson in heavy nuclei
 using the proton-nucleus reaction and
 to measure kaon pairs as well as the lepton pairs.
 By doing this, one can study not only the mass shift
but also the change of the leptonic vs hadronic branching ratio
$R = \Gamma(\phi \rightarrow e^+ e^- ) /\Gamma (\phi \rightarrow
K^+ K^- )$,
which is sensitive to the change of the $\phi$-mass  as well as
$K$-mass
 in medium.

 There are also on-going  heavy ion experiments at SPS (CERN) and
AGS (BNL)
 where  high density matter is likely to be formed.
 In particular, CERES/NA45 at CERN recently reported an
enhancement of the
 $e^+e^-$ pairs below the $\rho$ resonance, which is hard to be
explained
 by the conventional sources of the lepton pairs \cite{ceres}.
 Also, E859 at BNL-AGS reported a possible spectral change of the
 $\phi$-peak in $K^+K^-$ spectrum \cite{ags}.
 If these effects are real, the shoulder structure of the
spectrum
 expected by the mass shift of the vector mesons could be a
  possible explanation.

\section{Concluding remarks}

The spectral change of the elementary excitations in medium
 is an exciting new possibility in QCD.
 By studying such phenomenon, one can learn the structure
 of the hadrons and the QCD ground state at finite $(T, \rho$)
 simultaneously.   Some theoretical models
 predict that the  light vector mesons ($\rho$, $\omega$ and
$\phi$)
 are sensitive to the partial restoration of chiral symmetry
 in hot/dense medium.  These mesons are also experimentally good
 probes since they decay into lepton pairs which penetrate
 the hadronic medium without loosing much information.
  Thus, the lepton pair spectroscopy in QCD will tell us a
 lot about the detailed structure of the hot/dense matter, which
 is quite similar to the soft-mode spectroscopy
 by the photon and neutron scattering experiments in solid state
physics.
 The theoretical approaches to study the spectral changes
 are still in the  primitive stage and  new methods beyond
 QCD sum rules and naive effective lagrangian approaches
 are called for.

\end{document}